\documentclass[aps,twocolumn,english,superscriptaddress,citeautoscript,preprintnumbers,amsmath,amssymb,floatfix,footinbib]{revtex4-2}
\usepackage[breaklinks=true,colorlinks,citecolor=blue,linkcolor=blue,urlcolor=blue]{hyperref}
\usepackage{amsmath}
\usepackage{graphicx}
\usepackage{dcolumn}
\usepackage{float}
\usepackage[utf8]{inputenc}
\usepackage{color}
\usepackage[super]{nth}
\DeclareUnicodeCharacter{2212}{-}

\usepackage{mathtools}
\DeclarePairedDelimiter\bra{\langle}{\rvert}
\DeclarePairedDelimiter\ket{\lvert}{\rangle}

\usepackage[T1]{fontenc}

\begin{document}
	
\title{ Observation of quantum oscillations, linear magnetoresistance, and crystalline electric field effect in quasi-two-dimensional PrAgBi$_2$}

\author{Sudip Malick}
\email{sudip.malick@pg.edu.pl}
\author{Hanna Świątek}
\author{Michał J Winiarski}
\email{michal.winiarski@pg.edu.pl}
\author{Tomasz Klimczuk}
\email{tomasz.klimczuk@pg.edu.pl}
\affiliation{Faculty of Applied Physics and Mathematics, Gdansk University of Technology, Narutowicza 11/12, 80-233 Gdańsk, Poland}
\affiliation{Advanced Materials Center, Gdansk University of Technology, Narutowicza 11/12, 80-233 Gdańsk, Poland}

\begin{abstract}

We report the magnetic and magnetotransport properties with electronic band structure calculation of the Bi square net system PrAgBi$_2$. The magnetization and heat capacity data confirm the presence of a crystalline electric field (CEF) effect in PrAgBi$_2$. Analysis of the CEF effect using a multilevel energy scheme reveals that the ground state of PrAgBi$_2$ consists of five singlets and two doublets. The de Haas-van Alphen (dHvA) quantum oscillations data show a single frequency with a very small cyclotron effective mass of approximately 0.11 $m_e$. A nontrivial Berry phase is also observed from the quantum oscillations data. The magnetotransport data shows linear and unsaturated magnetoresistance, reaching up to 1060\% at 2 K and 9 T. Notably, there is a crossover from a weak-field quadratic dependence to a high-field linear dependence in the field-dependent magnetoresistance data. The crossover critical field $B^*$ follows the quadratic temperature dependence, indicating the existence of Dirac fermions. The band structure calculation shows several Dirac-like linear band dispersion near the Fermi level and a Dirac point close to the Fermi level, located at the Brillouin zone boundary. \textit{Ab inito} calculations allowed us to ascribe the observed dHvA oscillation frequency to a particular feature of the Fermi surface. Our study suggests layered PrAgBi$_2$ is a plausible candidate for hosting the CEF effect and Dirac fermion in the Bi square net.
	
\end{abstract}

\maketitle
\section{Introduction}	

Recently, quasi-two-dimensional (2D) pnictides have been receiving significant attention as they appear to host relativistic fermions. Such fermions are generated in these systems when two $p_{x,y}$ bands from Bi/Sb intersect and form linear energy dispersion, resulting in Dirac and Weyl states. Since electron transport in these compounds is driven by relativistic fermions, several quantum phenomena emerge, including linear magnetoresistance (MR), quantum oscillations, quantum Hall effect, large mobility, superconductivity and anisotropic transport properties \cite{Annual_Review, RSb2, CaBi2, PdBi2, CaMnBi2_PRB_2012, SrMnBi2_PRL_2011, EuMnBi2_SciAdv_2016, CaBi2_MW, LiBi, NaBi}. Consequently, these quantum materials show great potential for unique physical properties and practical applications.

$A$MnSb$_2$ and $A$MnBi$_2$ ($A$ = Ca, Sr, Ba, and Eu) are a well-studied series with quasi-2D structures containing Bi/Sb square sheets \cite{CaMnSb2_PRB_2017,CaMnSb2_ARPES_PRB_2021, SrMnSb2_2022, BaMnSb2_PNAS, BaMnSb2_PRB, EuMnSb2_PRB_2021, EuMnSb2_PRB_2022_Weyl, SrCaMnBi2_PRB_2014,SrCaMnBi2_SciRep_2014,SrCaMnBi2_NatCom_2016,CaSrMnBi2_PRB_2013, BaMnBi2_PRB_2016,BaMnBi2_SciRep_2017,EuMnBi2_PRB_2014}. It was found that these compounds crystallize mainly in tetragonal and orthorhombic structures. The different $A$ ions may lead to the formation of different crystal structures. (Ca/Sr)MnSb$_2$ crystallizes in the $Pnma$ space group \cite{CaMnSb2_PRB_2017,SrMnSb2_2022}, whereas CaMnBi$_2$ takes the $P4/nmm$ space group \cite{CaMnBi2_PRB_2012}; it changes to $I4/mmm$ in the case of (Sr/Ba/Eu)MnBi$_2$ \cite{SrMnBi2_PRL_2011,BaMnBi2_PRB_2016,EuMnBi2_PRB_2014}. Several measurements, such as electron transport, angle-resolved photoemission spectroscopy (ARPES) and first-principles calculations, confirm the existence of Dirac fermions in the Bi/Sb square nets \cite{Annual_Review,klemenz_role_2020,khoury_chemical_2021}. Some compounds show magnetism, breaking the time-reversal symmetry and leading to Weyl states. The physical properties of rare-earth ($R$) silver bismuthide and antimonide $R$AgBi$_2$ and $R$AgSb$_2$, which belong to the same family, were investigated a few decades ago \cite{RAgBi2_2003,RAgSb2}. Most compounds in this family exhibit long-range antiferromagnetic ordering competing with crystalline electric-field (CEF) effect at low temperatures. As an exception, the ground state of PrAgBi$_2$ was assumed to be non-magnetic, pending further investigations \cite{RAgBi2_2003}. These compounds have recently gained attention since they have been categorized as potential hosts for Dirac fermions. Linear MR has been observed in LaAgBi$_2$ and LaAgSb$_2$, which is ascribed to the quantum limit of the possible Dirac fermions \cite{LaAgBi2_PRB_2013,LaAgSb2_PRB_2012}. ARPES data suggest the existence of a Dirac cone in LaAgSb$_2$ along the $\Gamma$-$M$ high-symmetry line \cite{LaAgSb2_PRB_2016_ARPES}. Anisotropic electron transport and high mobility are observed in antiferromagnetic GdAgSb$_2$ \cite{GdAgSb2_2023} and SmAgSb$_2$ \cite{SmAgSb2}. Quantum oscillations study indicates a topological phase transition from nontrivial to trivial states as the system moves away from the antiferromagnetic states in SmAgSb$_2$ \cite{SmAgSb2}.

In this paper, we have grown high-quality single crystals of rare earth silver dibismuthide PrAgBi$_2$ and carried out a detailed analysis of the magnetic, thermodynamic, and magnetotransport data in conjunction with the band structure calculations. The magnetization and heat capacity data analyses confirm the presence of the CEF effect in PrAgBi$_2$. The de Haas-van Alphen (dHvA) oscillations data reveal a very low cyclotron effective mass and a nonzero Berry phase, which are consistent with the Dirac-like band dispersion observed in the band structure calculation. Additionally, magnetotransport data exhibits large unsaturated linear MR and a crossover from quadratic to linear field dependence. The Hall resistivity data suggest the presence of two types of carriers in PrAgBi$_2$.
The paper is organized as follows: The method of crystal growth, measurements, and band structure calculation is presented in Sec. \ref{Methods}. The crystal structure of PrAgBi$_2$ is briefly discussed in Sec. \ref{Structure}. Analysis of the CEF effect and dHvA oscillations is demonstrated in Sec. \ref{CEF_effect} and Sec. \ref{Oscillation}, respectively. Sec. \ref{Transport} and Sec. \ref{Hall_resistivity} discuss the magnetoresistance and Hall resistivity data, respectively. The band structure of PrAgBi$_2$ is presented in Sec. \ref{Band_structure}. Finally, we conclude our results in Sec. \ref{Summary}.

\section{Methods}
\label{Methods}

High quality single crystals of PrAgBi$_2$ and LaAgBi$_2$ were grown via self-flux method, using pieces of high purity Pr/La (99.9\%), Ag (99.99\%) and Bi (99.99\%) in a 1:1:10 ratio. The alumina crucible containing the elements was sealed in an evacuated, partially argon back-filled quartz tube. It was then heated to 1000$^\circ$C and held there for 12 h, followed by slow cooling at the rate of 3$^\circ$C/h. The tube was taken out of the furnace at 500$^\circ$C and centrifuged to separate the crystals from the flux. The crystal structure was determined by x-ray diffraction (XRD) using a Bruker D2 Phaser diffractometer with Cu K$\alpha$ radiation and a LynxEye XE-T detector. 
Physical properties were investigated using a Quantum Design Physical Property Measurement System with a vibrating sample magnetometer attachment for the magnetic measurements. The resistivity measurement employed a standard four probe technique with an applied current of 5 mA. Heat capacity was measured using a two-$\tau$ relaxation method. 
Electronic structure calculations were performed by means of the plane-wave basis Density Functional Theory (DFT) using the Quantum Espresso 7.2 package \cite{QE1,QE2,QE3}, employing the Perdew-Burke-Ernzerhof Generalized Gradient Approximation (PBE GGA) \cite{perdew_generalized_1996} of the exchange-correlation potential and the Projector-Augmented Wave (PAW) \cite{blochl_projector_1994} method in a fully-relativistic (spin-orbit coupling included) mode. Suitable PAW sets were taken from the PSLibrary \cite{dal_corso_pseudopotentials_2014}. The 4\textit{f} orbitals of Pr were treated as core states. The unit cell and structural parameters were relaxed using the Broyden-Fletcher-Goldfarb-Shanno (BFGS) to  method yielding a = 4.618 \AA, c = 10.444 \AA, Pr position (0, 0.5, 0.74839), Bi1 position (0, 0.5, 0.18396), with all other atomic parameters fixed by symmetry. Self-consistent calculations were completed on a 12$\times$12$\times$5 \textit{k}-point mesh with density and plane-wave cutoff energies set to 500 and 50 Ry, respectively. Fermi surface (FS) images were rendered using the FermiSurfer 2.4.0 program \cite{kawamura_fermisurfer:_2019}
The SKEAF program \cite{ROURKE2012324} was used to calculate the dHVA oscillation frequencies for each of the FS branches. In order to obtain a sufficiently dense \textit{k}-point grid, the FS file for SKEAF was calculated using an all-electron full potential linearized augmented plane wave (FP-LAPW) DFT code ELK 10.0.15, employing the PBE GGA. The correlation effects in the 4\textit{f} shell of Pr was treated using the DFT+U formalism in the fully-localized limit. The Hubbard U and exchange J parameters were set to 6.5 eV and 1.1 eV, respectively. The structural relaxation and self-consistent field calculations were completed using a 10x10x5 \textit{k}-point grid, while the Fermi surface was evaluated on a 51x51x31 grid in a non-SCF run. The resulting FS and a summary of dHvA calculations are presented in the Supplemental Material.

\section{Results and discussion}

\subsection{Crystal structure}
\label{Structure}

\begin{figure}
	\includegraphics[width=8.5cm]{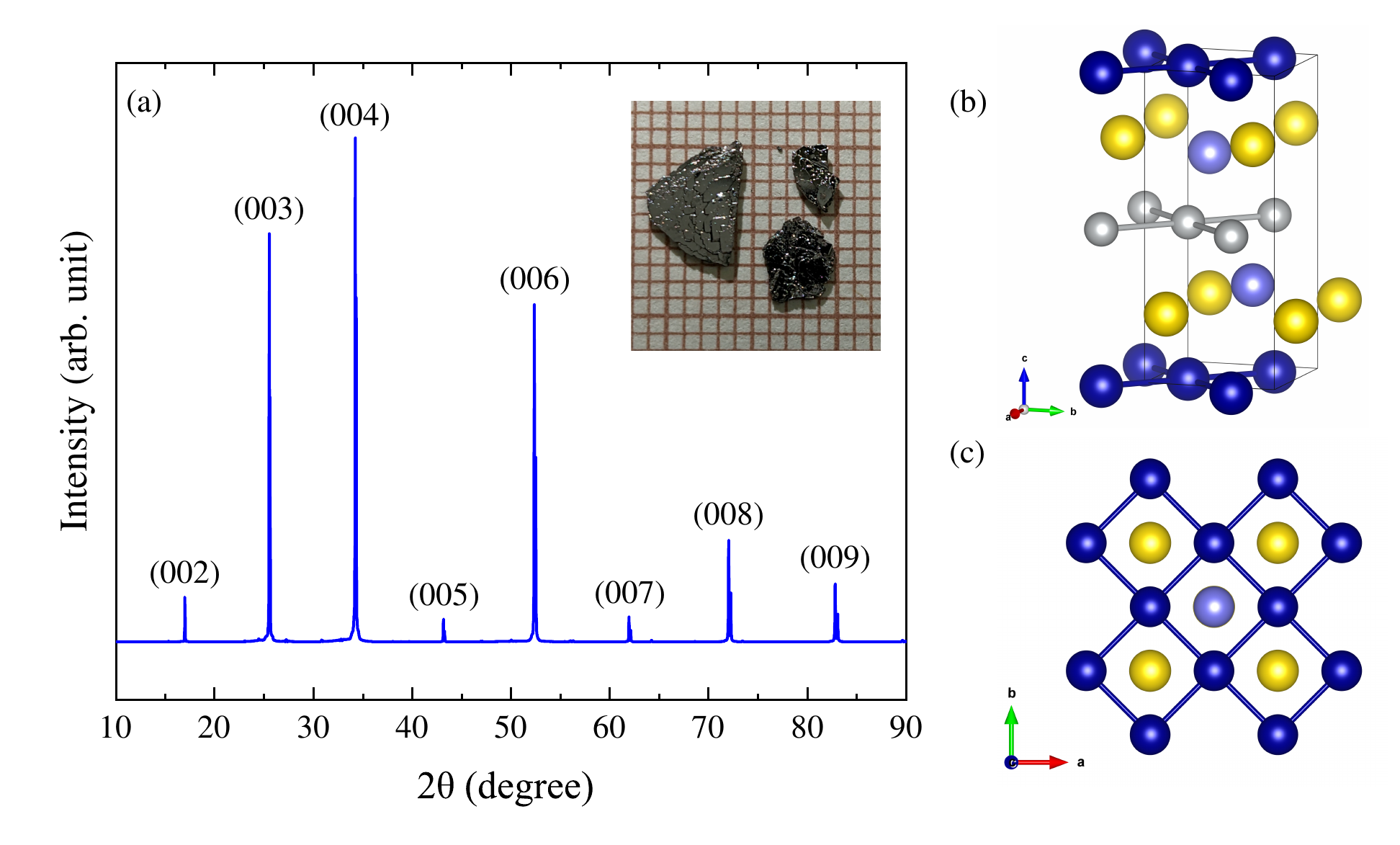}
	\caption{\label{xrd} (a) Single crystal XRD pattern of PrAgBi$_2$. Inset shows the optical image of single crystals (b) A schematic diagram of the crystal structure, with Pr atoms marked in yellow, Ag atoms in gray, and Bi1 and Bi2 in dark and light blue, respectively. (c) Bi square lattice seen along the $c$-axis.}
	
\end{figure}

The x-ray diffraction pattern obtained for a single crystal is presented in Fig \ref{xrd}(a). The compound crystallizes in a space group $P4/nmm$ (No. 129), adopting a layered tetragonal (ZrCuSi$_2$-type) structure commonly reported for 122 compounds. The observed XRD reflections match well with the expected (00$l$) positions, which confirms that the $c$-axis is perpendicular to the flat plane of the crystal. LeBail analysis of the data collected for powder samples (see Supplemental
Material \cite{SM}) leads to lattice parameters $a$ = 4.5078(2) \AA~ and $c$ = 10.4632(7) \AA~, which falls in between the previously reported values for the related compounds LaAgBi$_2$ and SmAgBi$_2$ \cite{RAgBi2_2003}, as expected due to the lanthanide contraction effect. The crystal structure is presented schematically in Fig \ref{xrd}(b). It is composed of alternating layers of Bi$^{-1}$ square net and [AgBi]$^{-2}$ slabs, separated by Pr$^{+3}$ ions. Fig \ref{xrd}(c) presents a closer look at the square Bi lattice viewed along the $c$-axis.

\subsection{Crystalline electric field}
\label{CEF_effect}

\begin{figure*}
	\centering
     
	\includegraphics[width=15cm, keepaspectratio]{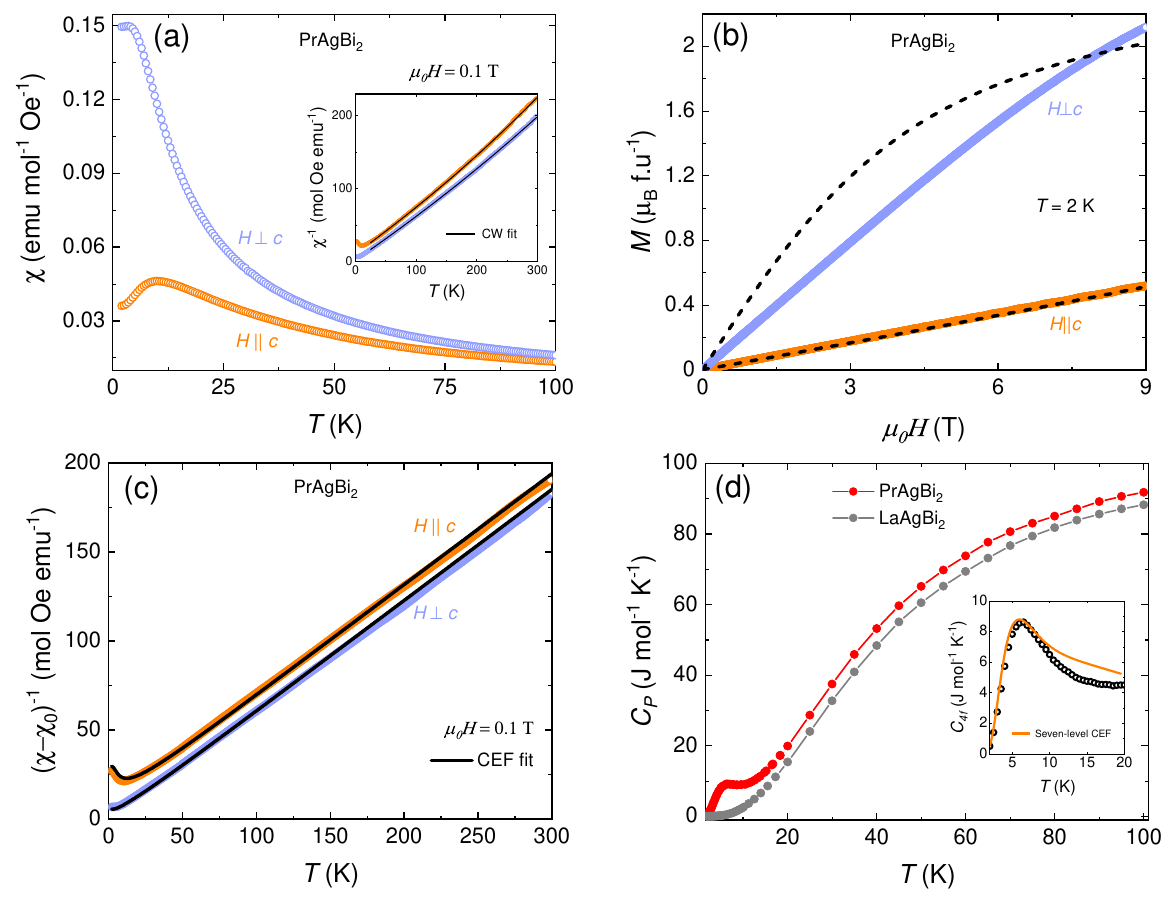}
	\caption{\label{MT_HC} (a) Temperature dependence of magnetic susceptibility of PrAgBi$_2$ along different crystallographic orientations measured at 0.1 T. The inset displays the inverse susceptibility as a function of temperature. The black line is the CW fit in the temperature range of 25–300 K. (b) Magnetic field-dependent isothermal magnetization measured at 2 K along two crystallographic orientations. Dotted lines are the calculated magnetization using CEF parameters. (c) The CEF fitting to the magnetic susceptibility for \textit{H}$\parallel$ \textit{c} and \textit{H}$\perp$ \textit{c} in the temperature range 2-300 K. (d) The temperature-dependent heat capacity of PrAgBi$_2$ and LaAgBi$_2$. The inset shows the $f$ electron contribution to the heat capacity in the temperature range of 2–20 K. The solid orange line is the Schottky contribution to the heat capacity calculated using CEF energy levels.}
	
\end{figure*}

\begin{table}

	\caption{\label{CW_fit}The estimated $\chi_0$, $\mu_{eff}$, and $\Theta_P$ from the modified Curie-Weiss fit to the inverse magnetic susceptibility data for different crystallographic directions of PrAgBi$_2$.} 
	\vskip .1cm
	\begin{tabular}{ccc}
		\hline
		\hline

		&~~~~~~~~~~\textit{H} $\parallel$ \textit{c}~~~~~~~~~~~~~& \textit{H} $\perp$ \textit{c}~~~~~~~~\\[1.5ex]
		\hline\\
		
		$\chi_0$ (10$^{-4}$ emu/mol)     &-7.86(3)  &-3.83(3)\\[1.5ex]
		$\mu_{eff}$ ($\mu_B$)           &3.7(1)   & 3.7(1)\\[1.5ex]
		$\Theta_P$ (K)                  &-18.2(1) 	&-3.8(1)\\[1.5ex]
		
		\hline
		\hline
		
	\end{tabular}
\end{table}

\begin{table*}

	\centering 
	\caption {\label{CEF_para}The estimated CEF parameters such as energy levels and the corresponding wave functions of PrAgBi$_2$}
	\vskip .1cm
	\begin{tabular}{c c c  c c c }
		\hline
		\hline \\[0.01ex]
		\multicolumn{6}{c}{CEF parameters}\\[1.5ex]
		\hline \\[0.01ex]
		~~~~~$B_2^0$ (K)~~~~~ &~~~~~~ $B_4^0$ (K) ~~~~~&~~~~~$B_4^4$ (K) ~~~~~~&~~~~~~$B_6^0$ (K)~~~~~&~~~~~~$B_6^4$ (K)~~~~~&~~~~~~$\lambda_i$ (mol/emu)~~~~ \\[1.5ex]
		0.91& 0.0045 & 0.14 & 0 & 0.0054&$\lambda_z$ = 2.7; $\lambda_{x,y}$ = -1.7        \\[1.5ex]
		
		\hline
		\hline\\
	\end{tabular}
	
\begin{tabular}{c c c c  c c c c cc}
\multicolumn{10}{c}{Energy levels and wave functions} \\[1.5ex]
\hline \\[0.01ex]
~~~E (K)~~~~&~~~~$\ket{+4}$~~~~&~~~~$\ket{+3}$~~~~&~~~~$\ket{+2}$~~~&~~~~$\ket{+1}$~~~~&~~~~
$\ket{0}$~~~~&~~~~$\ket{-1}$~~~~&~~~~$\ket{-2}$~~~~&~~~~$\ket{-3}$~~~~&~~~~$\ket{-4}$ \\[1.5ex]
		\hline\\
95.8 & 0.601 & 0 & 0 & 0 & 0.527 & 0 & 0 & 0 & 0.601 \\[1.5ex]
69.2 & -1/$\sqrt{2}$ & 0 & 0 & 0 & 0 & 0 & 0 & 0 & 1/$\sqrt{2}$\\[1.5ex]
54.6 &  0 & 0.577 & 0 & 0.409 & 0 & 0.409 & 0 & 0.577 & 0\\[1.5ex]
54.6 &  0 & -0.577& 0 & 0.409 & 0 & -0.409 & 0 & 0.577 & 0\\[1.5ex]
41.3 &  0 & 0 & -1/$\sqrt{2}$ & 0 & 0 & 0 & -1/$\sqrt{2}$ & 0 & 0     \\[1.5ex]
18.1 & 0 & 0 & -1/$\sqrt{2}$ & 0 & 0 & 0 & 1/$\sqrt{2}$ & 0 & 0\\[1.5ex]
12.9 &  0 & 0.409 & 0 & 0.577 & 0 & -0.577 & 0 & -0.409 & 0\\[1.5ex]
12.9 &  0 & 0.409 & 0 & -0.577 & 0 & -0.577 & 0 & 0.409 & 0 \\[1.5ex]
0.0 & 0.373 & 0 & 0 & 0 & -0.849 & 0 & 0 & 0 & 0.373 \\[1.5ex]
		
		\hline
		\hline
		
	\end{tabular}
\end{table*}

Fig. \ref{MT_HC}(a) displays the temperature-dependent magnetic susceptibility ($\chi = M/H$) data measured at $\mu_0H$ = 0.1 T for different crystallographic orientations of the PrAgBi$_2$ single-crystal. Below 12 K, the $\chi$(T) data exhibits a broad hump along $H \parallel c$, whereas susceptibility saturates along $H \perp c$ at low temperatures. Such temperature dependence of magnetic susceptibility indicates the existence of the CEF effect rather than a magnetic phase transition in PrAgBi$_2$ \cite{RAgBi2_2003,RAgSb2}. High-temperature susceptibility data follows modified Curie-Weiss (CW) behavior
\begin{equation}
	\chi(T) = \chi_0 + \dfrac{C}{T-\Theta_P}.
\end{equation} 

\noindent The inset of Fig. \ref{MT_HC}(a) illustrates the fitting of the inverse magnetic susceptibility to the modified CW equation in the temperature range of 25–300 K for  $H \parallel c$ and $H\perp c$. The obtained temperature independent susceptibility ($\chi_0$), effective moment ($\mu_{eff}$) and CW temperature ($\Theta_P$) are presented in Table \ref{CW_fit}. The so-obtained $\Theta_P$ is comparable to the previous report, and the estimated effective moment is close to the theoretical value (3.58 $\mu_B$) for Pr$^{3+}$ free-ions \cite{RAgBi2_2003}. The negative value of $\Theta_P$ indicates the predominance of antiferromagnetic interaction in the paramagnetic states. The isothermal magnetization $M(H)$ measured at 2 K for $H\parallel c$ and $H\perp c$ is shown in Fig. \ref{MT_HC}(b). Magnetization in PrAgBi$_2$ increases monotonically with increasing applied magnetic field without saturating even up to 9 T, as observed in several Pr-based CEF systems like PrAgSb$_2$, PrNi$_2$Cd$_{20}$, and PrPd$_2$Cd$_{20}$ \cite{RAgSb2,PrPd2Cd20}.


To address the anomaly in the low-temperature magnetic susceptibility data, we have analyzed the $\chi$($T$) data using a multilevel CEF scheme. PrAgBi$_2$ crystallizes in a tetragonal structure with point symmetry $D_{4h}$, thus for Pr$^{3+}$ (\textit{J} = 4) total multiplet (2\textit{J}+1) = 9, splits into two doublets and five singlets \cite{CEF_Levels, PrGaGe}. For such a system, the CEF Hamiltonian is defined as
\begin{equation}
\mathcal{H}_{CEF} = B_2^0O_2^0+B_4^0O_4^0+B_4^4O_4^4+B_6^0O_6^0+B_6^4O_6^4
\label{H_CEF}
\end{equation}

\noindent where $B_l^m$ and $O_l^m$ are the CEF parameters and the Stevens operators, respectively \cite{Stevens_1952,PT_CEF}. As per the CEF scheme, the temperature-dependent magnetic susceptibility is expressed as follows: 

\begin{equation}
\begin{split}
\chi_{CEF}^i
= &\frac{N_A(g_J\mu_B)^2}{Z} \bigg[ \sum_{n}\beta|\bra{n} J_i \ket{n} |^2 e^{-\beta E_n}  \\
&+\sum_{n\ne m}|\bra{m} J_i \ket{n}|^2 \frac{e^{-\beta E_n}-e^{-\beta E_m}}{E_m-E_n} \bigg].
\end{split}
\label{MT_CEF_Eq}
\end{equation} 

\noindent Here $N_A$ and $g_J$ are the Avogadro constant and Land\'{e} factor, respectively. $Z=\sum_{n}e^{-\beta E_n}$, where $\beta$ = 1/$k_BT$ and $\ket{n}$ is the $n$th eigenfunction with eigenvalue $E_n$. $J_i$ ($i$ = $x$, $y$, and $z$) is the component of angular momentum \cite{CeIr3Ge7}. The first term of the Eq. \ref{MT_CEF_Eq} represents the Curie contribution to the susceptibility, and the second is the Van Vleck susceptibility. The overall magnetic susceptibility, including molecular field ($\lambda_i$) contribution, is calculated as

\begin{equation}
	\label{CEF_MS}
	(\chi_i-\chi_0)^{-1} = (\chi_{CEF}^i)^{-1}-\lambda_i.
\end{equation}

\noindent The calculated temperature-dependent inverse magnetic susceptibilities data for different crystallographic directions are in good agreement with the experimentally observed data, as shown in Fig. \ref{MT_HC}(c). The estimated CEF parameters and corresponding energy levels and wave functions from the calculation are presented in Table \ref{CEF_para}. To ensure the consistency of the analysis, a CEF parameter $B_2^0$ is directly calculated from the Curie-Weiss temperatures for different crystal orientations using the formula \cite{WANG1971_CEF}

\begin{equation}
\Theta_P^{ab}-\Theta_P^c = \frac{3}{10}B_2^0(2J-1) (2J+3),
\label{Theta}
\end{equation}

\noindent which yields $B_2^0$ = 0.62, close to the value estimated from the CEF model fit.

Additionally, we use the following Hamiltonian to estimate the isothermal magnetization data using the CEF model

\begin{equation}
	\mathcal{H} = \mathcal{H}_{CEF}-g_j \mu_BJ_i(H+\lambda_i M_i),
	\label{M_CEF}
\end{equation}

\noindent where $\mathcal{H}_{CEF}$ is the CEF Hamiltonian mentioned in Eq. \ref{H_CEF}. The second and third terms of the above expression represent the Zeeman and molecular field contributions to the total Hamiltonian, respectively. The magnetization ($M_i$) along different crystallographic directions can be calculated using the expression
 
\begin{equation}
M_i= \frac{g_J\mu_B}{Z} \sum_{n}|\bra{n} J_i \ket{n} | e^{-\beta E_n}.  \\
\label{Mag}
\end{equation} 

\noindent The eigenvalue and associated eigenfunction of the above expression are derived by diagonalizing the entire Hamiltonian described in Eq. \ref{M_CEF}. The calculated magnetization at 2 K using CEF parameters for $H\parallel c$ and $H\perp c$ is approximately replicates the experimental data as displayed in Fig. \ref{MT_HC}(b). However, this computation precisely recognizes the anisotropy of magnetization.

Fig. \ref{MT_HC}(d) shows the temperature-dependent heat capacity $C_P$($T$) for PrAgBi$_2$ and the nonmagnetic reference LaAgBi$_2$. The heat capacity data exhibits a rather broad anomaly around 7 K, which may be associated with the Schottky-type anomaly resulting from the splitting of energy levels due to CEF effect, as seen in the similar compound PrAgSb$_2$ \cite{RAgSb2}. The Schottky contribution to the heat capacity for the multilevel system can be expressed as:

\begin{equation}
	\begin{split}
		C_{Sch}(T)
		=&\left( \dfrac{R}{T^2}\right) \bigg[  \sum_{i} g_ie^{-\Delta_i/T}\sum_{i}g_i{\Delta_i}^2e^{-\Delta_i/T}\\
		& - \left( \sum_{i}g_i\Delta_ie^{-\Delta_i/T}\right)^2\bigg]  \left( \sum_{i}g_ie^{-\Delta_i/T}\right)^{-2},
	\end{split}
\end{equation} 

\noindent where $R$ is the molar gas constant and $g_i$ is the degeneracy of the \textit{i}th state with $\Delta_i$ energy gap splitting \cite{gopal_specific}. The estimated Schottky heat capacity using the same CEF energy levels as given in Table \ref{CEF_para} fairly reproduces the 4$f$ contribution to the specific heat ($C_{4f}$), calculated using the formula $C_{4f}$ = $C_P$(PrAgBi$_2$) - $C_P$(LaAgBi$_2$), as displayed in the inset of Fig. \ref{MT_HC}(d). This finding further validates the estimated CEF energy levels in PrAgBi$_2$.

\subsection{Quantum oscillations}
\label{Oscillation}

\begin{figure*}

	\centering
	\includegraphics[width=16cm, keepaspectratio]{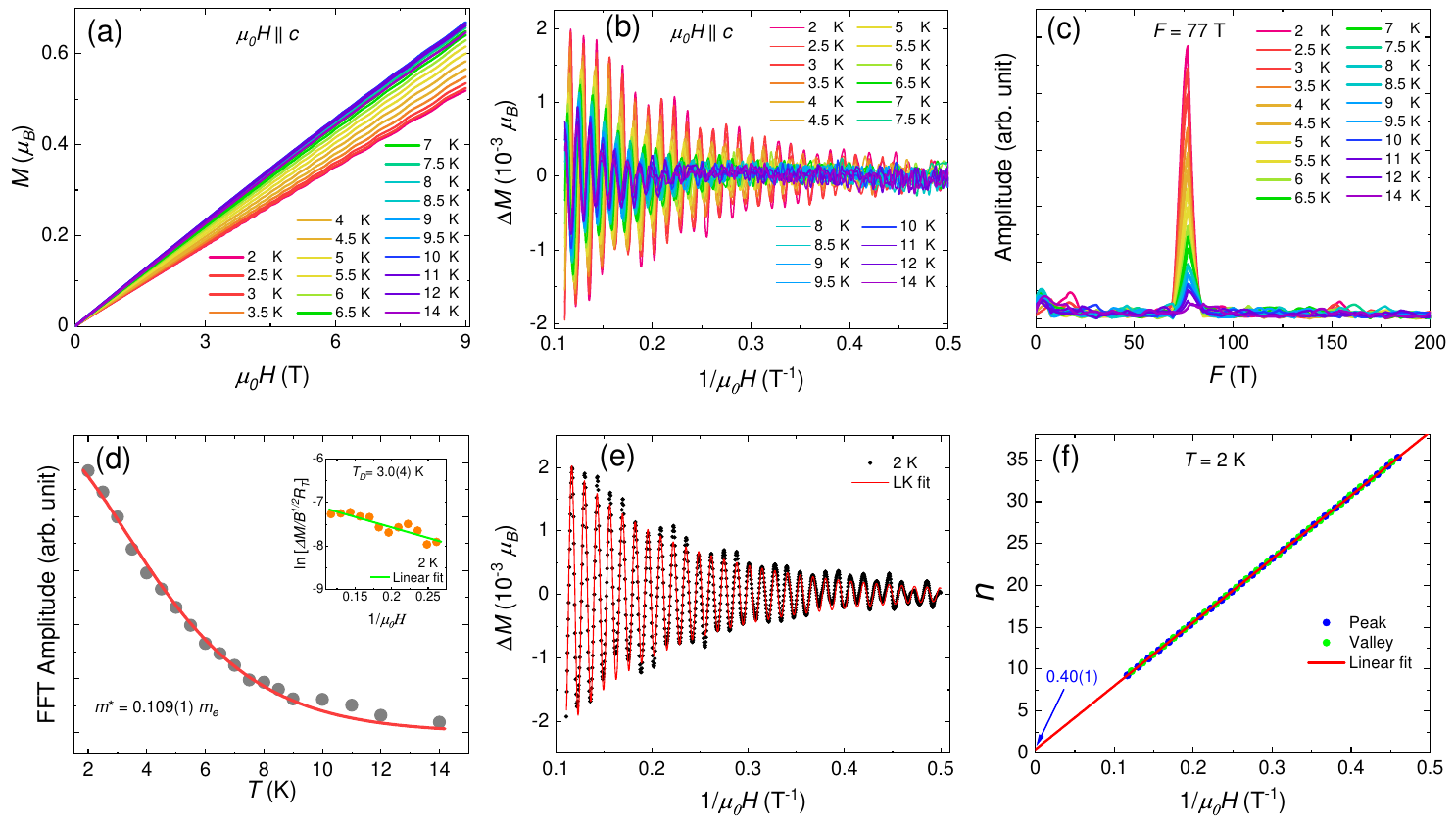}
	\caption {\label{QO}(a) Field-dependent magnetization for various temperatures along $H||c$. (b) The oscillatory component of the magnetization data as a function of 1/$\mu_0H$. (c) The fast Fourier transform spectra of dHvA oscillations data for various temperatures. (d) Temperature dependence of FFT amplitude. The solid red line is a fit to the temperature-dependent term $R_T$ of the LK equation. The inset shows the Dingle plot. (e) Fitting of dHvA data with the LK equation at 2 K. (f) LL fan diagram: The peaks ($n$+1/4) and valleys ($n$-1/4) are assigned blue and green dots, respectively.}
\end{figure*}

The dHvA quantum oscillations are observed in the field-dependent isothermal magnetization ($M$) data along $H\parallel c$ in the high field region and temperature up to 14 K, as depicted in Fig. \ref{QO}(a). The oscillatory component of the magnetization $\Delta M$ is extracted by subtracting a polynomial background resulting in prominent oscillations, as shown in Fig. \ref{QO}(b). The oscillation amplitude of the magnetization can be expressed using the Lifshitz-Kosevich (LK) formula \cite{shoenberg}

\begin{equation}
	\Delta M \propto -B^{1/2}R_TR_D\text{sin}\left[ 2\pi\left( \dfrac{F}{B}-\gamma-\delta)\right)\right]  
\end{equation}

\noindent where $R_T$ is the thermal damping factor, which is defined as $R_T= \dfrac{\alpha T\mu/B}{\text{sinh}(\alpha T\mu/B)}$, $\alpha$ = 14.69 T/K and $\mu = m^*/m_e$. $m^*$ is the effective cyclotron mass. The term $R_D = \text{exp}(\alpha T_D\mu/B)$ stands for the Dingle damping factor with Dingle temperature $T_D$, which is connected to the quantum scattering lifetime ($\tau_q$) through the expression $\tau_q = \hbar/(2\pi k_BT_D$). The phase factor $\gamma$ is directly related to the Berry phase ($\Phi_B$) through the expression $\gamma = 1/2-\Phi_B/2\pi$. The factor $\delta$ represents the additional phase shift, which is 0 for 2D systems and $\pm$1/8 for 3D systems  \cite{ZrSiSe}. Fig. \ref{QO}(b) shows the fast Fourier transform (FFT) spectra of dHvA oscillations data at various temperatures, indicating a fundamental frequency of $F$ = 77 T. The effective cyclotron mass is calculated by fitting the temperature-dependent FFT amplitudes to $R_T$, which yields $m^*$ $\simeq$0.11 $m_e$ as shown in Fig. \ref{QO}(d). The obtained effective mass is comparable to several isostructural compounds such as LaAgBi$_2$ \cite{LaAgBi2_PRB_2013}, GdAgSb$_2$ \cite{GdAgSb2_2023}, SmAgSb$_2$ \cite{SmAgSb2}, and \textit{A}MnBi$_2$ \cite{SrMnBi2_PRL_2011, CaMnBi2_PRB_2012}. The orthogonal cross-sectional area ($A$) corresponding to the frequency $F$ can be calculated using the Onsager equation $F = (\hbar/2\pi e)A$, resulting in $A$ = 0.73 nm$^{-2}$. Assuming circular Fermi-surface cross-section Fermi wave vector ($k_F$) and velocity ($v_F$) are estimated to be 0.48 nm$^{-1}$ and 0.51$\times$10$^6$ ms$^{-2}$, respectively, using the expression $k_F = \sqrt{2eF/\hbar}$ and $v_F = \hbar k_F/m^*$. Further Dingle temperature is calculated from the field-dependent oscillatory part of the LK equation as presented in the inset of Fig. \ref{QO}(d). The obtained $T_D$ = 3 K and corresponding quantum scattering lifetime $\tau_Q$ = 4.05$\times$10$^{-13}$ s and quantum mobility $\mu_Q = e\tau_Q/m^*$ = 6534 cm$^2$V$^{-1}$s$^{-1}$. All the estimated parameters are summarized in Table \ref{dHvA_para}.

\begin{table*}
	
	\centering 
	\caption {\label{dHvA_para} Parameters obtained from the dHvA quantum oscillations.}
	\vskip .1cm
	\begin{tabular}{ccccccccc}
		\hline
		\hline \\[0.01ex]
		$F$(T)~~~~& $A$(nm$^{-2}$)~~~~& $k_f$(nm$^{-1}$)~~~~& $v_f$ (10$^6$ms$^{-1}$)~~~~& $m^*/m_e$~~~~& $T_D(K)$~~~~& $\tau_Q$ (10$^{-13}$s)~~~~& $\mu_Q$ (cm$^2$V$^{-1}$s$^{-1}$)~~~~& $\Phi_B$ ($\delta$ =0)\\[1.5ex]
		
		\hline\\[0.01ex]
    77& 0.73& 0.48& 0.51& 0.11& 3& 4.05& 6534& 0.8$\pi$\\[1.5ex]
		\hline
		\hline \\[0.01ex]
		
	\end{tabular}
	
\end{table*}

To get an estimation of the Berry phase, dHvA oscillations data is fitted to the LK equation as shown in Fig. \ref{QO}(e). To reduce the dependency of too many fitting parameters, we use the effective mass and Dingle temperature that we obtained from previous analysis. The estimated value of the Berry phase is 0.84$\pi$. Further, the Berry phase is also estimated from the Landau level (LL) fan diagram to get a more reliable value. The LL index is assigned as $n$+1/4 and $n$-1/4 against the peaks and valleys, respectively, for the $\Delta M$ data at 2 K, as shown in Fig. \ref{QO}(f). The slope of the linear fit of the LL indices indicates the oscillation frequency equal to 75.9 T, which is very close to that obtained from the FFT, validating the constructed LL fan diagram. The intercept $n_0$ = 0.40, and the corresponding Berry phase can be calculated using the expression $\Phi_B = 2\pi(n_0+\delta)$ \cite{TiSb2, La3MgBi5, LuSb2}. For $\delta$ = 0, $\Phi_B = 0.8\pi$, which is comparable to that estimated for the LK fit. The obtained values of $\Phi_B$ taking $\delta$ = $\pm$1/8 are 1.05$\pi$ and 0.55$\pi$, respectively. The nonzero Berry phase, which serves as an indicator of the topological nature of a system, implies the possible existence of Dirac fermions in PrAgBi$_2$ \cite{SrMnBi2_PRL_2011, CaMnBi2_PRB_2012, SmAgSb2}.

\subsection{Linear magnetoresistance}

\begin{figure*}
	
	\centering
	\includegraphics[width=16cm, keepaspectratio]{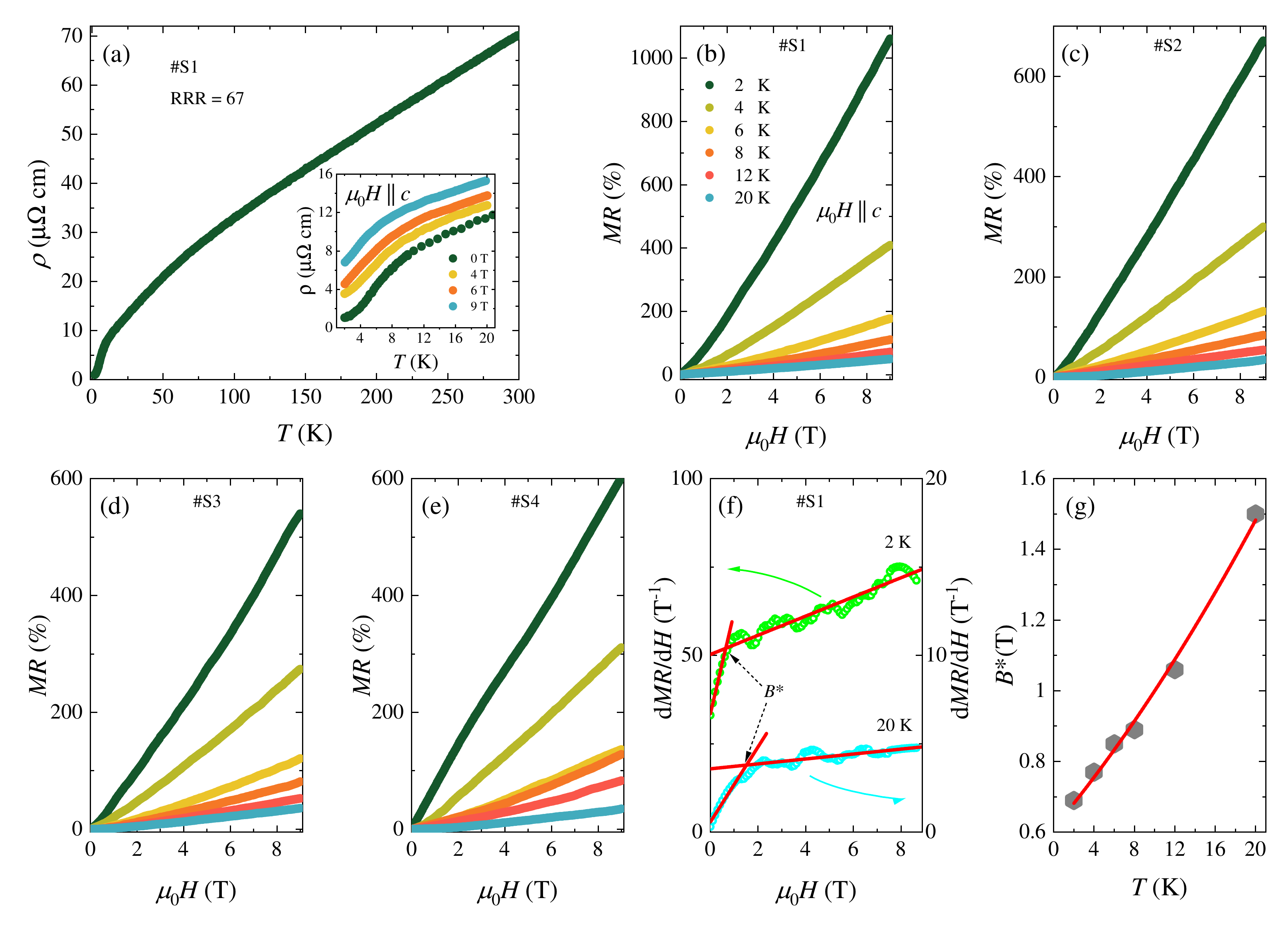}
	\caption {\label{MR}(a) Electrical resistivity as a function of temperature in the absence of an external magnetic field. Inset shows the resistivity in the temperature range 2-20 K for the applied magnetic field $\mu_0 H$ = 0, 4, 6, 9 T. (b)-(d) Field-dependent magnetoresistance measured up to 9 T on different single crystals at $T$ = 2, 4, 6, 8, 12, and 20 K. (f) First-order derivative of magnetoresistance with respect to the applied field at 2 K and 20 K. (g) The estimated critical field as a function of temperature. The red line is the fitting of the data with the expression of $B^*(T)$ as described in the text.}
\end{figure*}
\label{Transport}

\begin{figure}
	\includegraphics[width=8.5cm]{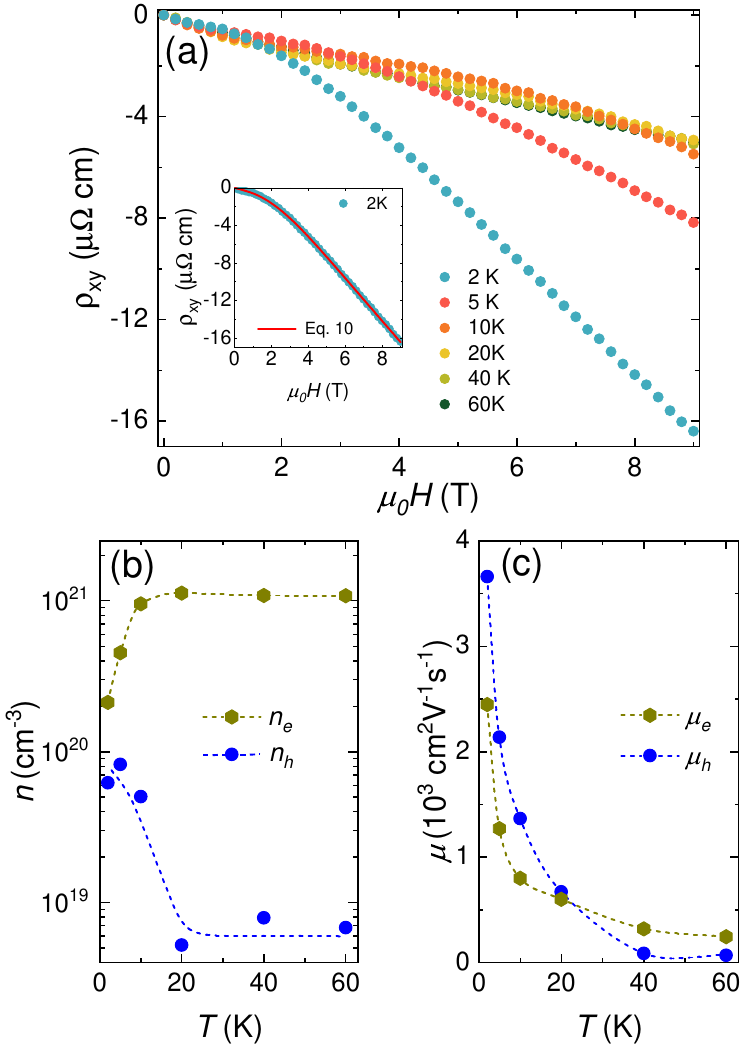}
	\caption{\label{Hall} (a) Hall resistivity as a function of magnetic field of PrAgBi$_2$ measured in the temperature range 2-60 K. The inset shows the Hall resistivity at 2K, fitted with the two-band model (Eq. \ref{RhoXY}). The temperature dependence of carrier concentrations ($n_e$ \& $n_h$) and mobilities ($\mu_e$ \& $\mu_h$) is shown in (b) and (c), respectively.}
	
\end{figure}

The electrical resistivity  $\rho$(\textit{T}) of single-crystalline PrAgBi$_2$ as a function of temperature for current flowing in the \textit{ab}-plane is shown in Fig. \ref{MR}(a). The electrical resistivity decreases monotonically with decreasing temperature, showing metallic behavior. No pronounced anomaly is seen; however, around 10 K, a hump-like feature is observed in the $\rho$(\textit{T}) data, consistent with the previous report \cite{RAgBi2_2003}. The broad peak in  $\rho$(\textit{T}) could be associated with CEF effect seen in the temperature-dependent magnetic susceptibility and heat capacity data. At 2 K, a significant decrease in the resistivity is observed, resulting in a large residual resistivity ratio (RRR). The electrical transport measurements were carried out on four single crystals obtained from different crystal growths. The RRR value ranges from 50 to 67, indicating that the as-grown crystals are of excellent quality and the results are reproducible. The inset of Fig. \ref{MR}(a) shows the  $\rho$(\textit{T}) in the temperature range of 2–20 K under a few constant applied magnetic fields along the $c$-axis. The resistivity is enhanced significantly without changing its feature as the applied field increase, indicating large magnetoresistance in PrAgBi$_2$.

\begin{figure*}
	
	\centering
	\includegraphics[width=16cm, keepaspectratio]{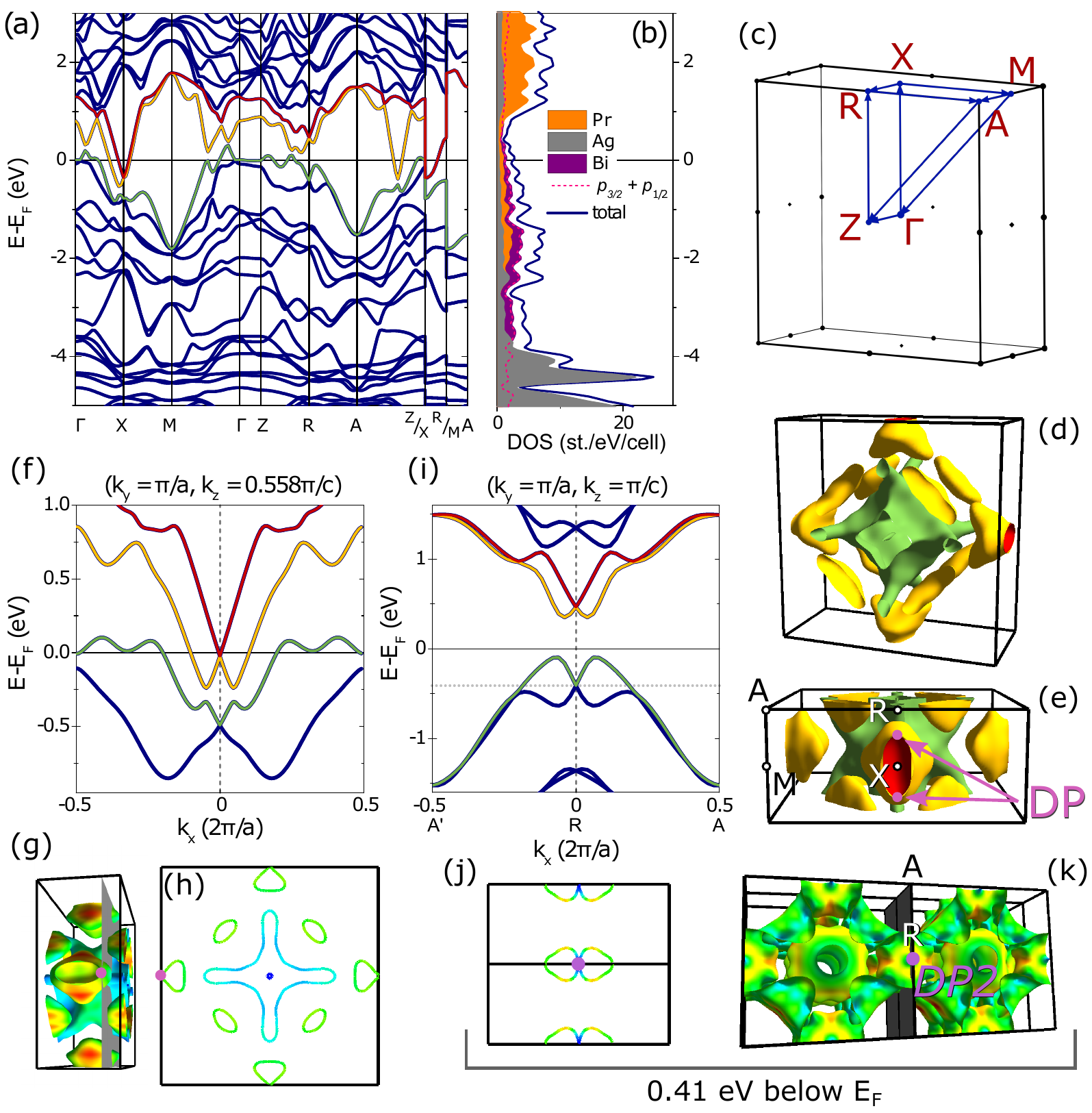}
	\caption {\label{dft} Band structure (a) and density of states (b) of PrAgBi$_2$. Three bands crossing the Fermi level are highlighted in green, yellow and red. (c) High symmetry points of the first Brillouin zone (BZ). Panels (d) and (e) present two projections of the calculated Fermi surface of PrAgBi$_2$. Dirac points lying close to the Fermi level are labeled in pink, high-symmetry points - in black and white. Panel (f) shows the band dispersion around the Dirac point positioned at $k$ = (0.0, 0.5, 0.279) in crystal coordinates, which lies at the Fermi level. Panels (g) and (h) show two sections of the FS colored according to the modulus of the Fermi velocity. In panel (g) the section cuts through the Dirac point (one of the four DPs at the plane is highlighted in pink). Panel (i) shows the band dispersion along the A$^\prime$-R-A line. A symmetry protected crossing (DP2, purple) is found at the R point, 0.41 eV below the Fermi level. The band shown in green forms one of the branches of the Fermi surface, which is responsible for the observed dHvA oscillations (see SM \cite{SM}). Panels (j) and (k) show the Fermi surface for the Fermi energy shifted downwards by 0.41 eV, at which energy the second Dirac point (DP2) is found.}
\end{figure*}

The field-dependent magnetoresistance measurements were conducted on four different crystals in the temperature range 2-20 K for \textit{H}$\parallel$\textit{c} direction and current in the \textit{ab}-plane, as illustrated in Figs. \ref{MR}(b)-\ref{MR}(e). Large, nonsaturating linear MR is observed in all crystals. MR reaches 1060\% at 2 K and 9 T in sample S1, whereas other samples display a relatively low MR. The crystal quality strongly influences the value of MR as observed in the case of LaAgBi$_2$ \cite{LaAgBi2_LargeMR}. In addition, a slight misalignment of the sample plane with the applied magnetic field may also lead to variation in the MR, as the Fermi surface of these series of compounds is highly anisotropic \cite{LaAgBi2_PRB_2013}. Nevertheless, all crystals display a similar field and temperature dependence. The MR shows a strong temperature dependence; as the temperature increases, the MR drops significantly. Besides the linear MR in the high field, PrAgBi$_2$ exhibits a semiclassical quadratic field dependency in the weak field regime. The quadratic nature becomes more pronounced at higher temperatures. The field-dependent MR at low temperatures in PrAgBi$_2$ manifests a significant deviation from semiclassical transport and a violation of Kohler's scaling (see SM \cite{SM}), implying the involvement of relativistic fermions, multiband, or quantum effects in magnetotransport \cite{LaAgSb2_PRB_2012,LaAgBi2_PRB_2013, MR_Metals, Kohler, Extended_Kohler's}. Recently, linear and unsaturated MR has been observed in several quasi-2D system featuring a Dirac band dispersion, including BaFe$_2$As$_2$ \cite{BaFe2As2}, Bi/Sb square net LaAgBi$_2$ \cite{LaAgBi2_PRB_2013}, LaAgSb$_2$ \cite{LaAgSb2_PRB_2012}, and CaMnBi$_2$ \cite{CaMnBi2_PRB_2012, CaMnBi2_APL_2012}. Such unusual MR in these compounds has been interpreted using Abrikosov’s theorem of quantum limit, in which all the carriers (electrons or holes) in the Dirac states are confined to the zeroth Landau level. In such scenario, the Fermi energy $E_F$ and thermal fluctuations $k_BT$ must be less than the energy level splitting between the zeroth and first LLs, denoted by $\Delta_{LL} =\pm v_F\sqrt{2\hbar eB}$. Dirac fermions have a high Fermi velocity; thus, the splitting energy for relativistic fermions is much higher than that of conventional fermions in the parabolic bands, where LL splitting energy $\Delta$ = $\hbar eB/m^*$. As a result, a quantum limit can be realized in moderately applied magnetic fields in topological materials \cite{BaFe2As2, Abrikosov, Graphene, Annual_Review_TS}. It can be anticipated that the Dirac-like dispersive band in PrAgBi$_2$ may lead to large, linear, and nonsaturating MR as observed in isostructural systems like LaAgBi$_2$ \cite{LaAgBi2_PRB_2013} and CaMnBi$_2$ \cite{CaMnBi2_PRB_2012}. Moreover, a crossover from the quadratic to linear dependence in MR is pointed out as the critical field $B^*$ in the $\frac{d\text{MR}}{d\textit{H}}$ vs. $\mu_0 H$ data as shown in Fig. \ref{MR}(f). In the low-field region, the first derivative of MR follows linear dependence, reflecting semi-classical MR behavior, Whereas in the higher field it has a much-reduced slope, implying a dominating linear response with a small quadratic factor in the field-dependent MR. The temperature dependence of the critical field, as presented in Fig. \ref{MR}(g), is well described by the expression $B^*(T) = \dfrac{1}{2e\hbar v_F^2}(E_F+k_BT)^2$ \cite{BaFe2As2,LaAgBi2_PRB_2013, CaMnBi2_PRB_2012}, where Fermi energy $E_F$ = 2.9 meV and Fermi velocity $v_F$ = 0.64 $\times$ 10$^6$ ms$^{-1}$. This further indicates the influence of Dirac fermions on the magnetotransport properties of PrAgBi$_2$, consistent with our band structure calculation discussed below.

\subsection{Hall resistivity}
\label{Hall_resistivity}

The field dependence of Hall resistivity $\rho_{xy}$ measured in the 2-60 K temperature range is presented in Fig. \ref{Hall}(a). Hall resistivity exhibits a negative slope and nonlinear behavior at low temperatures; as the temperature increases, $\rho_{xy}$ becomes linear. These observations suggest that the transport in PrAgBi$_2$ involves two types of carriers, with electrons being the predominant type. The semi-classical two-band model was used to estimate the concentrations and mobilities of the carriers, where Hall resistivity is defined as,
\begin{equation}
    \rho_{xy}= \frac{\mu_0H}{e}\dfrac{(n_h\mu_h^2-n_e\mu_e^2)+(n_h-n_e)\mu_e^2\mu_h^2(\mu_0H)^2}{(n_h\mu_h+n_e\mu_e)^2+(n_h-n_e)^2\mu_e^2\mu_h^2(\mu_0H)^2},
    \label{RhoXY}
\end{equation}

\noindent where, $n_e$ and $n_h$ ($\mu_e$ and $\mu_h$) are the carrier concentration (mobility) of the electrons and holes, respectively. Figures. \ref{Hall}(c) and \ref{Hall}(d), depict the temperature variation of carrier concentrations and mobilities estimated by fitting Eq. \ref{RhoXY} to the Hall resistivity data. At low temperatures, the estimated electron (hole) concentration $\sim10^{20}$ cm$^{-3}$ ($\sim10^{19}$ cm$^{-3}$) and the mobility $\sim10^3$ cm$^2$V$^{-1}$s$^{-1}$ are comparable to several topological materials \cite{LaAgSb2_PRB_2012, YbMnSb2,SmAgSb2}. The electron concentration increases to $\sim10^{21}$ cm$^{-3}$ as the temperature increases, whereas hole concentration decreases to $\sim10^{18}$ cm$^{-3}$. A significant drop in mobility is observed in both cases as the temperature increases.

\subsection{Electronic band structure}
\label{Band_structure}
Fig. \ref{dft} presents the results of electronic structure calculations of PrAgBi$_2$. As expected, the band structure (Fig. \ref{dft}(a)) is remarkably similar to the isoelectronic LaAgBi$_2$ \cite{LaAgBi2_LargeMR} and LaAuBi$_2$ \cite{seibel_structure_2015, yu_first-principles_2022}. The density of states at the Fermi level, $DOS(E_F)$, is mostly contributed by Bi 6\textit{p} states as shown in Fig. \ref{dft}(b). This suggests that the electronic properties of PrAgBi$_2$ derive from the square Bi$^{-1}$ network, in agreement with chemical bonding considerations \cite{Annual_Review,klemenz_role_2020,khoury_chemical_2021}. The calculated Fermi surface (FS), shown in Fig. \ref{dft}(d,e), consists of three branches (marked with green, yellow, and red). Two of the FS branches (marked in red and yellow) touch at a high symmetry line \textit{W} spanning from the X point at the zone boundary to the R point at the zone edge, forming a Dirac point (DP) at $k \approx$ (0.0, 0.5, 0.279) in crystal coordinates (see  Fig. \ref{dft}(f-h)), protected by the non-symmorphic \textit{n} glide symmetry operation. Band dispersions around the DP in the vicinity of $E_F$ are plotted in Fig. \ref{dft}(f). The calculated Fermi velocity in the vicinity of the DP are slightly above $0.5\times10^6$ ms$^{-1}$ (Fig. \ref{dft}(g)), in a reasonable agreement with the value estimated from the magnetoresistance analysis. 

The calculation of the dHvA oscillation frequencies (see the SM\cite{SM}) indicates that the observed oscillation with frequency $F$ = 77 T can be attributed to a set of orbits contained in the branch of the FS shown in green in Fig. \ref{dft} (d,e). This band has a nontrivial character, forming a Dirac point located at 0.41 eV below the $E_F$ at the \textit{R} point of the BZ [see Figs. \ref{dft}(i-k)]. Such observation is consistent with the value of the Berry phase estimated from the dHvA oscillation data, which suggests the existence of Dirac fermions in PrAgBi$_2$.

The two bands marked in yellow and red in  Fig. \ref{dft}(a,f,i) have a linear dispersion around the $E_F$ and their crossing lies very close to the Fermi level. In case of the remaining band, the Dirac point is located below the Fermi level and the dispersion relation is nonlinear around $E_F$. Thus the results of electronic structure calculations allow us to assume that linear MR and observed dHvA oscillations result from two separate parts of the Fermi surface - the former is associated with the Dirac bands crossing at the Fermi level at $k$ = (0.0, 0.5, 0.279), while the latter stems from tubular features of one of the FS branches (see the SM \cite{SM} Fig. S4), which forms a Dirac point at the R point of the BZ, positioned 0.41 eV below the Fermi level.

\section{Conclusion}
\label{Summary}

In conclusion, we have carried out a systematic analysis of magnetic and magnetotransport properties, together with the band structure calculation of high-quality single crystals of PrAgBi$_2$. The analysis of magnetization and heat capacity data confirms the crystalline electric field effect in the absence of any magnetic phase transition in PrAgBi$_2$. The ground-state energy level is determined to be divided into two doublets and five singlets. The de Haas-van Alphen quantum oscillations data shows a nonzero Berry phase and a very small cyclotron effective mass, indicating nontrivial topological state. A linear and unsaturated magnetoresistance in PrAgBi$_2$ is observed in the magnetotransport measurements. Interestingly, the field-dependent magnetoresistance data show a crossover from a weak-field quadratic dependency to a high-field linear dependence. The crossover critical field $B^*$ exhibits a quadratic temperature dependence, implying the splitting of linear energy dispersion in PrAgBi$_2$. The band structure calculation reveals bands with linear dispersion and a Dirac point located at the BZ boundary, which is likely responsible for the observed linear MR, while the dHvA oscillation was ascribed to particular nontrivial band that forms a protected crossing ca. 0.41 eV below the Fermi level. Therefore, our investigation indicates the existence of Dirac fermions in the quasi-two-dimensional CEF system PrAgBi$_2$.

\section{Acknowledgment} 
Research performed at Gdansk Tech. was supported by the National Science Center (Poland), Project No.
2022/45/B/ST5/03916. S. M. thanks Daloo Ram for valuable discussions.

\bibliography{PrAgBi2}

\end{document}